\documentclass[a4paper,11pt]{article}
\pdfoutput=1 

\usepackage{jheppub} 

\usepackage[T1]{fontenc} 
\usepackage[italian,english]{babel}
\usepackage{hyperref}
\usepackage{ifpdf}
\usepackage{subfigure}
\usepackage{tikz}
\usetikzlibrary{arrows,shapes}
\usetikzlibrary{trees}
\usetikzlibrary{matrix,arrows} 				
\usetikzlibrary{positioning}				
\usetikzlibrary{calc,through}				
\usetikzlibrary{decorations.pathreplacing}  
\usepackage{pgffor}							

\usetikzlibrary{decorations.pathmorphing}	
\usetikzlibrary{decorations.markings}
\tikzset{
    vector/.style={decorate, decoration={snake}, draw},
    fermion/.style={draw=black, postaction={decorate}}, 
    scalar/.style={dashed,draw=black, postaction={decorate}}}
\tikzstyle{block} = [draw, rectangle, 
    minimum height=3em, minimum width=6em]

\usepackage{amssymb}
\usepackage{amsfonts}
\usepackage{epsf}
\usepackage{rotating}
\usepackage{graphicx}
\usepackage{amsmath}
\usepackage{fancyhdr}
\usepackage{lineno}
\usepackage{babel}
\usepackage{graphics}
\usepackage{pstricks}
\usepackage{color}
\usepackage{multirow}

\newcommand{\nn}{\nonumber}
\newcommand{\lsim}{\mathrel{\mathop{\kern 0pt \rlap
  {\raise.2ex\hbox{$<$}}}
  \lower.9ex\hbox{\kern-.190em $\sim$}}}
\newcommand{\gsim}{\mathrel{\mathop{\kern 0pt \rlap
  {\raise.2ex\hbox{$>$}}}
  \lower.9ex\hbox{\kern-.190em $\sim$}}}

\newcommand{\be}{\begin{equation}}
\newcommand{\ee}{\end{equation}}
\newcommand{\bea}{\begin{eqnarray}}
\newcommand{\eea}{\end{eqnarray}}

\def\gev{\ensuremath{\mathrm{\,Ge\kern -0.1em V\,}}}
\def\tev{\ensuremath{\mathrm{\,Te\kern -0.1em V\,}}}

\def\phid{\ensuremath{\phi_\text{\tiny DM}}\,}
\def\phidst{\ensuremath{\phi^*_\text{\tiny DM}}\,}
\def\Yd{Y_\text{\tiny DM}}
\def\qd{q_\text{\tiny DM}}
\def\invfb   {\ensuremath{\text{\,fb}^{-1}}\,}

\title{\boldmath  Implications of right-handed neutrinos in $B-L$ extended standard model with scalar dark matter}

\author[a]{Priyotosh Bandyopadhyay,}
\author[b]{Eung Jin Chun}
\author[c]{and Rusa Mandal}

\affiliation[a]{Indian Institute of Technology Hyderabad, Kandi,  Sangareddy-502287, Telengana, India}
\affiliation[b]{Korea Institute for Advanced Study, Seoul 130-722, Korea}

\affiliation[c]{The 
	Institute  of Mathematical Sciences, Taramani, Chennai 600113, India \\ and \\ Homi Bhabha National Institute Training School Complex, \\ Anushakti Nagar, Mumbai 400085, India}

\emailAdd{bpriyo@iith.ac.in} 
\emailAdd{ejchun@kias.re.kr} 
\emailAdd{rusam@imsc.res.in}

\preprint{ IITH-PH-0001/17

\hspace*{11.27cm} IMSc/2017/07/05}

\abstract{ We investigate the Standard Model (SM) with a $U(1)_{B-L}$ gauge extension where a $B-L$ charged scalar is a viable dark matter (DM) candidate. The dominant annihilation process, for the DM particle is through the $B-L$ symmetry breaking scalar to right-handed neutrino pair. We exploit the effect of decay and inverse decay of the right-handed neutrino in thermal relic abundance of the DM. 
Depending on the values of the decay rate, the DM relic density can be significantly 
different from what is obtained in the standard calculation assuming the right-handed neutrino is in thermal equilibrium and there appear different regions of the parameter space  satisfying the observed DM  relic density.
For a DM mass less than $\mathcal{O}(\tev)$, the direct detection experiments impose a competitive bound on the mass of the $U(1)_{B-L}$ gauge boson $Z^\prime$ with the collider experiments. Utilizing the non-observation of the displaced vertices arising from the right-handed neutrino decays, bound on the mass of $Z^\prime$ has been obtained at present and higher luminosities at the LHC with 14 TeV center of mass energy where an integrated luminosity of 100\,fb$^{-1}$ is sufficient to probe $m_{Z'} \sim 5.5$ TeV.}

\begin{document}
\maketitle
\flushbottom
\section{Introduction}

The indirect astrophysical evidence of the existence of  missing mass in form of a matter, called dark matter (DM), and the confirmation of the existence of tiny neutrino mass through neutrino oscillation are  the two major motivations to look for possible extension of the standard model (SM). According to the Planck data~\cite{Ade:2015xua}, about a fourth of the energy density of the Universe consists of  DM. However, in the absence of any direct observation, we are still in the darkness about the nature of DM. From the last three decades, plethora of candidates have been imagined as a DM particle in literature. Among them one of the most popular choices is a weakly interacting and massive particle (WIMP) whose mass lies in the GeV to TeV range with typically weak interactions. A WIMP pair annihilation to the SM particles provides a natural mechanism to produce the WIMP at the early Universe and can also explain the observed DM density in current Universe. As the mass range lies within the range GeV to TeV, these particles are accessible in current or future colliders as well as in different direct and indirect detection experiments of DM.

An enormous number of extensions of the SM are studied where the DM particle may have an integer or a half integer spin. Stabilization of a DM occurs naturally in the supersymmetric models where the lightest supersymmetric particle acts as a viable DM candidate. However in plenty of beyond standard model (BSM) extensions, an ad hoc discrete symmetry is imposed to forbid the decay of a DM particle. The validity of the assumption that the discrete symmetry 
remains unbroken due to gravitational effects at the Planck scale suffers from a suspicion~\cite{Boucenna:2012rc,Mambrini:2015sia}. This problem is eluded by identifying it to some high scale physics which is beyond the scope of the model under consideration.

In this paper, we consider an extension of the SM with a $U(1)_{B-L}$ gauge group. One possible way to cancel the gauge anomaly is by including three right-handed neutrinos in the theory. Thus, the model can naturally incorporate the light neutrino masses through Type-I seesaw mechanism. Various ideas to incorporate a DM candidate in the $U(1)_{B-L}$ context have been explored  in literature~\cite{Basak:2013cga}. An attractive option is to introduce a SM-singlet but $U(1)_{B-L}$-charged scalar particle which is stabilized by judicious choice of the $B-L$ quantum numbers. 
The main purpose of this work is  to revisit the case where the DM pair annihilates to right-handed neutrino pair through the $B-L$ symmetry breaking scalar and investigate the effect of right-handed neutrino decay and inverse decay in the thermal history of the DM particle. The effect of right-handed neutrino decay in a context of a supersymmetric $U(1)^\prime$ extension of the SM was studied in Ref.~\cite{Bandyopadhyay:2011qm}. We will see that  such an effect plays an important role in keeping the DM in thermal equilibrium and extending the allowed parameter space satisfying the observed relic density. We present that the measurements of spin-independent (SI) cross-section of DM-nuclei scattering in direct detection experiments especially XENON1T~\cite{Aprile:2017iyp} can impose bound on the mass of $B-L$ gauge boson $Z^\prime$ superior to the collider limits. The decay of a right-handed neutrino also provides very interesting and rich phenomenology from the collider aspects. The presence of displaced vertex arising from right-handed neutrino decay allows us the impose indirect limit on the mass of $Z^\prime$ at the LHC with current as well as higher integrated luminosities.

The paper is organized as follows. First we briefly describe main features of the model in Sec.~\ref{sec:model}. In Sec.~\ref{sec:relic} right-handed neutrino decay and inverse decay effects in the relic density of DM particles are discussed. Section~\ref{sec:directD} deals with the direct detection experiment limits. In Sec.~\ref{sec:RHnu} we explore the LHC signature of the pair production of right-handed neutrinos from the decay of $Z^\prime$  and also the decay of right-handed neutrino into the SM particles. Finally we conclude in Sec.~\ref{sec:conclusion} with discussion.

\section{The model}
\label{sec:model}

In this section we briefly discuss the basic setup of the model used in our work. We consider the extension of the SM with a gauged $U(1)_{B-L}$ symmetry. Apart from the SM particles, the model contains; a $U(1)_{B-L}$ gauge boson $Z^\prime$, three right-handed neutrinos $N_i$ to cancel the $B-L$ gauge anomaly, two SM singlet $B-L$ charged complex scalar $S$ and $\phid$ where $\phid$ is the would-be dark matter candidate. The interaction terms in the Lagrangian due to the new particles are given by,
\begin{align}
	\label{lag}
\mathcal{L}_\text{NP}&=  - m_S^2 |S|^2- \frac{1}{2} \lambda_{SH} {|S|}^2 |\Phi|^2  - \lambda_{S} (S^\dagger S)^2  - \lambda_{N_i} S \bar{N_i^c} N_i- y_{ij} \bar{L_i} \Phi^\dagger N_j \nn \\
& - m_D^2 |\phid\!|^2- \frac{1}{2} \lambda_{D H} {|\phid\!|}^2 |\Phi|^2- \frac{1}{2} \lambda_{D S} {|\phid\!|}^2 |S|^2  - \lambda_{D} (\phid^{\!\dagger} \phid)^2.
\end{align}
The $\Phi$ and $L_i$ are the usual SM Higgs and lepton $SU(2)_L$ doublets, respectively. In Table.~\ref{Table}, we show the $B-L$ charge assigned for all the SM and BSM particles.  
After the $B-L$ symmetry breaking through the vacuum expectation value (vev) of the scalar $S$, mass term for the $B-L$ gauge boson $Z^\prime$ as well as the Majorana mass for the neutrino $N_i$ are generated. The masses for light SM neutrinos are generated through the usual Type-I seesaw mechanism:
\begin{equation}
 {\cal M}^\nu_{ij} = y_{ik} y_{jk} {\langle \Phi \rangle^2 \over m_{N_k}}.
\end{equation}
For the low-scale Type-I seesaw with $m_{N_k} \sim$ TeV, one typically needs Yukawa couplings as small as $y_{ik}\sim 10^{-6} $ to generate neutrino mass scale around 0.1 eV. While lepton flavour violation (LFV) induced by such tiny couplings can hardly appear in low-energy observables, 
LFV signatures may appear at the LHC through the right-handed neutrino production and decays  \cite{LFVpheno}. On the other hand,  in the case of inverse seesaw where $y_{ik}$ can be of order one,
the induced LFV could be observed in various low-energy processes depending on the models \cite{LFVlow} and also interestingly in  exotic Higgs decays \cite{Arganda:2004bz,pbej, LFVconst}.

\begin{table}[h]
	\centering
	\begin{tabular}{ |c| c |c |c| c|c|c|c |}
		\hline \hline
		 & $Q$ & $u^c,d^c$ &  $L$ & $e^c$ & $N_i$ & $S$ & $\phid$ \\ \hline
		$B-L $ & $1/3$ & $-1/3$ & $-1$ & $1$ &$-1$ & $2$ &$\qd$ \\
		\hline
		\hline 
	\end{tabular}
	\caption{$B-L$ charges for all the particles present in the model.	}  \label{Table}
\end{table}

Being a scalar DM candidate, $\phid$ is forbidden to get VEV and mix with the symmetry breaking fields $S$ and $\Phi$. Thus the scalar potential relevant for the gauge symmetry breaking is,
\begin{align}
V(\Phi,S)= m_H^2 |\Phi|^2 + m_S^2 |S|^2 + \frac{1}{2}\lambda_H (\Phi^\dagger \Phi)^2  +\lambda_{S} (S^\dagger S)^2 + \frac{1}{2}\lambda_{SH} {|S|}^2 |\Phi|^2.
\end{align}

After spontaneous symmetry breaking (SSB), $\Phi= (H^+,H)^T$ with $H=(v+h)/\sqrt{2}$ and $S=(v^\prime + S_0+i S^\prime)/\sqrt{2}$ the scalar mass matrix is given by,
\begin{align}
\mathcal{M}(h,S_0)= \begin{pmatrix}
v^2 \lambda_H~~~~~\frac{1}{2}v\,v^\prime \lambda_{SH} \cr
\frac{1}{2}v\,v^\prime \lambda_{SH} ~~~2{v^\prime}^2 \lambda_S \cr
\end{pmatrix},
\end{align}

where we have used the following minimization conditions,
\begin{align}
\frac{\partial V}{\partial \Phi}\bigg|_{v,v^\prime}&= 0 \Longrightarrow
m_H^2= \frac{1}{4} \left(2v^2 \lambda_H+{v^\prime}^2 \lambda_{SH}\right), \\
\frac{\partial V}{\partial S}\bigg|_{v,v^\prime}&= 0 \Longrightarrow
m_S^2= \frac{1}{4} \left(4{v^\prime}^2 \lambda_S+ v^2 \lambda_{SH}\right).
\end{align} 
Allowing mixing between the two neutral components of $\Phi$ and $S$, we can write
\begin{align}
\begin{pmatrix}
\Phi \cr
S \cr
\end{pmatrix}= 
\begin{pmatrix}
\text{cos}\,\alpha~~~~~\text{sin}\, \alpha \cr
-\text{sin}\,\alpha~ ~~~\text{cos}\, \alpha \cr
\end{pmatrix} \begin{pmatrix}
h \cr
S_0 \cr
\end{pmatrix},
\end{align}
where the mixing angle is defined as
\begin{align}
\label{eq:alpha}
\text{tan}\,2\alpha = \frac{v\,v^\prime \lambda_{SH}}{v^2 \lambda_H-{2v^\prime}^2 \lambda_S}.
\end{align}
The two mass eigenstates of the scalar bosons are,
\begin{align}
m_{h,S_0}^2=\frac{1}{2}\bigg(v^2 \lambda_H+{2v^\prime}^2 \lambda_S\mp \sqrt{\big(v^2 \lambda_H-{2v^\prime}^2 \lambda_S\big)^2+ v^2 {v^\prime}^2 \lambda_{SH}^2} \bigg).
\end{align}
In view of the current bound on the mixing parameter $\alpha$ from the measurements of the Higgs boson properties at the LHC, we assume almost vanishing mixing between the two scalars implying cos$\alpha\simeq 1$ through out our analysis.

The SSB of $B-L$ symmetry provides mass term for the $B-L$ gauge boson $Z^\prime$ given by
\begin{align}
m_{Z^\prime}= 2 \,g_{BL}v^\prime, 
\end{align}
where $g_{BL}$ is the $B-L$ gauge coupling constant. At tree level, we assume no kinetic mixing between $U(1)_{B-L}$ and $U(1)_Y$ gauge bosons and hence $Z^\prime$ and the SM $Z$ boson do not mix with each other.

The right-handed neutrinos $N_i$ acquire mass after the SSB of $B-L$ which is $m_{N_i}\sim \sqrt{2} v^\prime \lambda_{N_i}$. We assume that only one of the three $N_i$'s is lighter than the DM candidate and thus relevant for our discussion. For simplicity, we denote the  mass eigenstate of the lightest right-handed neutrino by $N$\footnote{Left-handed neutrinos ($\nu_i$) and right-handed neutrinos ($N_i$) mix in their mass basis and we get three light neutrinos and three heavy neutrinos. } and the corresponding Yukawa coupling with the SM Higgs and lepton $SU(2)_L$ doublet as $y_N$. It turns out that, to satisfy the observed relic abundance, $y_N$ cannot be large enough to produce the SM neutrino mass larger than $\sim 10^{-3}\,$eV. That is, the contribution of $N$ to the SM neutrino mass matrix is negligible and hence only the other two heavy right-handed neutrinos are relevant to explain the observed neutrino masses and mixing. We note that in general the Yukawa matrix $y_{ij}$ contains off diagonal elements however its detailed texture is not relevant for the purpose of this paper.


The complex scalar \phid does not acquire vev and has a $B-L$ charge $\qd$. As discussed in Ref.~\cite{Rodejohann:2015lca}, some particular choices of $\qd$ i.e., $\qd\not =\pm 2 n$ for $n\in \mathbb{Z}$ and $n\le 4$, forbid  the \phid to decay and hence it can be a dark matter candidate without invoking any extra symmetry in the theory. We chose $\qd=1/2$ for most of our analysis. As we discuss later that in our case, the dominant process for the DM annihilation cross-section is through $s$-channel $S_0$ exchange, the charge of the DM candidate \phid does not affect the results obtained in Sec.~\ref{sec:relic}. However the direct detection bounds depend on the $B-L$ charge of \phid and will be addressed in Sec.~\ref{sec:directD}.
The mass term for $\phid$ receives contribution from both the EW and $B-L$ symmetry breaking given by
\begin{align}
m_{DM}^2= m_D^2 + \frac{1}{4}\lambda_{DH}v^2 + \frac{1}{4}\lambda_{DS}{v^\prime}^2.
\end{align}

It can be seen from Eq.~\eqref{lag} that the Yukawa interaction of the right-handed neutrino allows it to decay to SM particles via the mixing with the SM neutrinos proportional to $y_N$ and below we quote the expressions of the decay widths of $N$ to three possible channels $h\nu$, $\ell^\pm W^\mp$ and $Z\nu$, respectively, where we assume cos$\,\alpha\simeq1$.
\begin{align}
\label{eq:Ndecayh}
\Gamma(N\to h \nu)&=\Gamma(N\to h \bar{\nu})= \frac{y_N^2 m_N}{64\pi} \left(1- \frac{m_h^2}{m_N^2}\right)^2, \\ 
\Gamma(N\to \ell^- W^+)&=\Gamma(N\to \ell^+ W^-)= \frac{y_N^2 m_N}{32\pi} \left(1- \frac{m_W^2}{m_N^2}\right)^2 \left(1+ 2 \frac{m_W^2}{m_N^2}\right), \\ 
\label{eq:NdecayZ}
\Gamma(N\to Z \nu )&=\Gamma(N\to Z \bar{\nu})= \frac{y_N^2 m_N}{64\pi} \left(1- \frac{m_Z^2}{m_N^2}\right)^2 \left(1+ 2 \frac{m_Z^2}{m_N^2}\right).
\end{align}

\section{Relic density of the scalar dark matter}
\label{sec:relic}

\begin{figure}[t]
	\begin{center}
		\includegraphics[width=0.8\linewidth]{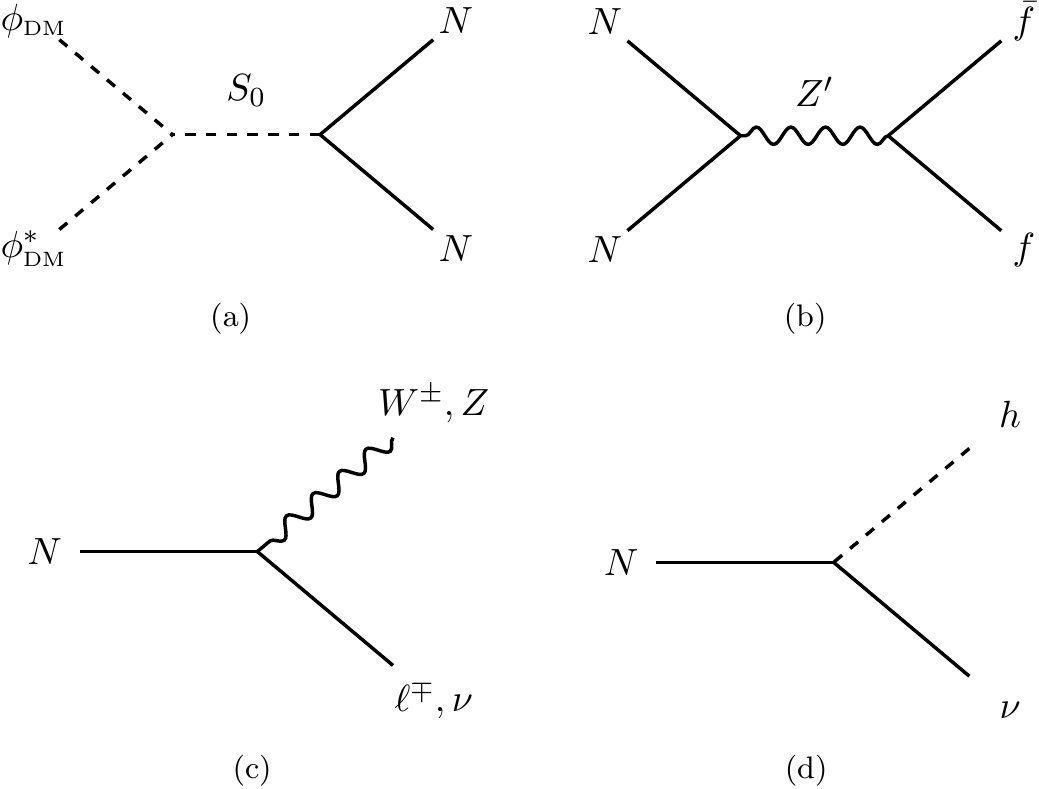}
		\caption{The Feynman diagrams for the DM particle \phid annihilation to right-handed neutrino $N$ pair (a), annihilation of $N$ to SM fermion anti-fermion $f\bar{f}$ (b) and the decay of $N$ to SM particles (c), (d) are shown. }\label{dia:relic}
	\end{center}
\end{figure}

In this section we discuss the thermal relic abundance of the scalar DM candidate \phid. The \phid can annihilate through three different interactions: the SM Higgs $h$ portal, the $B-L$ scalar $S_0$ portal, and the $B-L$ gauge boson $Z^\prime$ portal. The simplistic scenario of the Higgs portal is very strongly constrained~\cite{He:2016mls} by the SI cross-section measurements at direct detection experiments. Alternatively, as the bound on the mass of a new gauge boson from collider experiments is currently $\ge 2.8 \tev$ \cite{z'bnd}, to produce the observed relic abundance through $Z^\prime$ portal, one needs very heavy DM particle where the DM pair can annihilate via resonant production of $Z^\prime$. In this paper, we are interested in low mass ($\le \mathcal{O}(\tev)$) DM particle where the DM annihilates through the $B-L$ scalar $S_0$, predominantly. It can be seen from Eq.~\eqref{lag} that the $B-L$ scalar can couple directly to only one SM particle i.e., the Higgs doublet $\Phi$, and thus interacts to the SM fermions through mixing. To satisfy the current LHC bounds on the measurements of Higgs boson properties, the mixing angle $\alpha$ (defined in Eq.~\eqref{eq:alpha}) should be small. Hence we assume a tiny $\alpha$ of the $\mathcal{O}({10}^{-3})$ for our analysis. It can be easily understood that due to the presence of tree level coupling of $S$ and the right-handed neutrino $N$, the dominant annihilation cross-section of the DM candidate \phid is
through the process $\phid \phidst \to N N$ as shown in Fig.~\ref{dia:relic}(a). As long as $N$ is in thermal equilibrium for long enough time, the relic density can be estimated by calculating the thermal average of the annihilation process mentioned above. However, $N$ interacts through the heavy gauge boson $Z^\prime$ and also through the tiny Yukawa coupling $y_N$ and hence the interaction may not be very weak to keep $N$ in thermal equilibrium through the process of freeze-out of \phid. Thus, to study the thermal history of \phid through the annihilation in Fig.~\ref{dia:relic}(a), one also has to consider the evolution of $N$ determined by its annihilation (Fig.~\ref{dia:relic}(b)) and decay (Figs.~\ref{dia:relic}(c) and \ref{dia:relic}(d)).

We start with the coupled Boltzmann equations written in terms of the variable $Y_i\equiv n_i/s$, describing the actual number of particle $i$ per comoving volume, where $n_i$ being the number density, $s$ is the entropy density of the Universe, and the variable $x\equiv m_{DM}/T$ as
\begin{align}
\label{eq:dYDM}
\frac{d\Yd}{dx}&= -\frac{1}{x^2} \frac{s(m_{DM})}{H(m_{DM})} \langle \sigma v \rangle_{\phid\phidst\to NN} \left(\Yd^2 -Y_N^2\right), \\
\label{eq:dYN}
\frac{d Y_N}{dx}&= \frac{1}{x^2} \frac{s(m_{DM})}{H(m_{DM})} \langle \sigma v \rangle_{\phid \phidst \to NN} \left(\Yd^2 -Y_N^2\right) \nn \\ 
&-\frac{1}{x^2} \frac{s(m_{DM})}{H(m_{DM})} \langle \sigma v \rangle_{NN\to f\bar{f}} \left(Y_N^2 -{Y_N^{\text{eq}}}^2\right) -\frac{\Gamma}{H(m_{DM})} x \left(Y_N -Y_N^{\text{eq}}\right).
\end{align}

\noindent
The entropy density $s$ and Hubble parameter $H$ at the DM mass is
$$
s(m_{DM})= \frac{2 \pi^2 }{45} g_*\, m_{DM}^3, \quad H(m_{DM})= \frac{\pi}{\sqrt{90}} \frac{\sqrt{g_*}}{M^r_{pl}} m_{DM}^2, $$  where $M^r_{pl}= 2.44\times {10}^{18}\gev$ is the reduced Planck mass and $Y_N^{\text{eq}}$ is the equilibrium number density of right-handed neutrino $N$ given by
\begin{align}
\label{eq:YN_eq}
Y_N^{\text{eq}}\equiv\frac{n_N^{\text{eq}}}{s} &=\frac{45}{2\pi^4} \sqrt{\frac{\pi}{8}}\left( \frac{g}{g_*}\right) \left({\frac{m_N}{T}}\right)^{3/2} e^{-\frac{m_N}{T}} \nn \\
&\simeq 0.145 \left( \frac{2}{100}\right)  \left( \frac{m_N}{m_{DM}}\right)^{3/2} x^{3/2} e^{-\frac{m_N}{m_{DM}} x}.
\end{align}
Here in the last line of Eq.~\eqref{eq:YN_eq} we use the effective number of relativistic degrees of freedom $g_*\simeq100$ and the internal degrees of freedom $g=2$ for the complex scalar DM candidate \phid.
The first terms on the right-hand side of Eqs.~\eqref{eq:dYDM} and \eqref{eq:dYN} denote the forward and backward reactions of \phid \phidst to $NN$ through $s$-channel $S_0$ exchange shown in Fig.~\ref{dia:relic}(a). The second term on the right-hand side of Eq.~\eqref{eq:dYN} refers to the forward and backward reactions of $NN$ annihilation to the SM fermion pairs $f\bar{f}$ through the s-channel $Z^\prime$ exchange (Fig.~\ref{dia:relic}(b)) and the third term describes the decay and the inverse decay of $N$ shown in Fig.~\ref{dia:relic}(c) and (d) where $\Gamma$ being the total decay width of $N$.

The DM candidate \phid remains in thermal equilibrium through the interaction of $N$.  The right-handed neutrino $N$ annihilates to the SM fermion pair via the process $NN \to f \bar{f}$ (Fig.~\ref{dia:relic}(b)). As the current limit from LHC on $m_{Z^\prime}\ge 2.8 \tev$, the mentioned annihilation cross-section is very suppressed and thus the right-handed neutrinos freezes out earlier than the DM particle \phid. As a consequence, the DM particles are overproduced.

\begin{figure}[t]
	\begin{center}
		\mbox{\hskip -20 pt \subfigure[]{\includegraphics[width=0.5\linewidth]{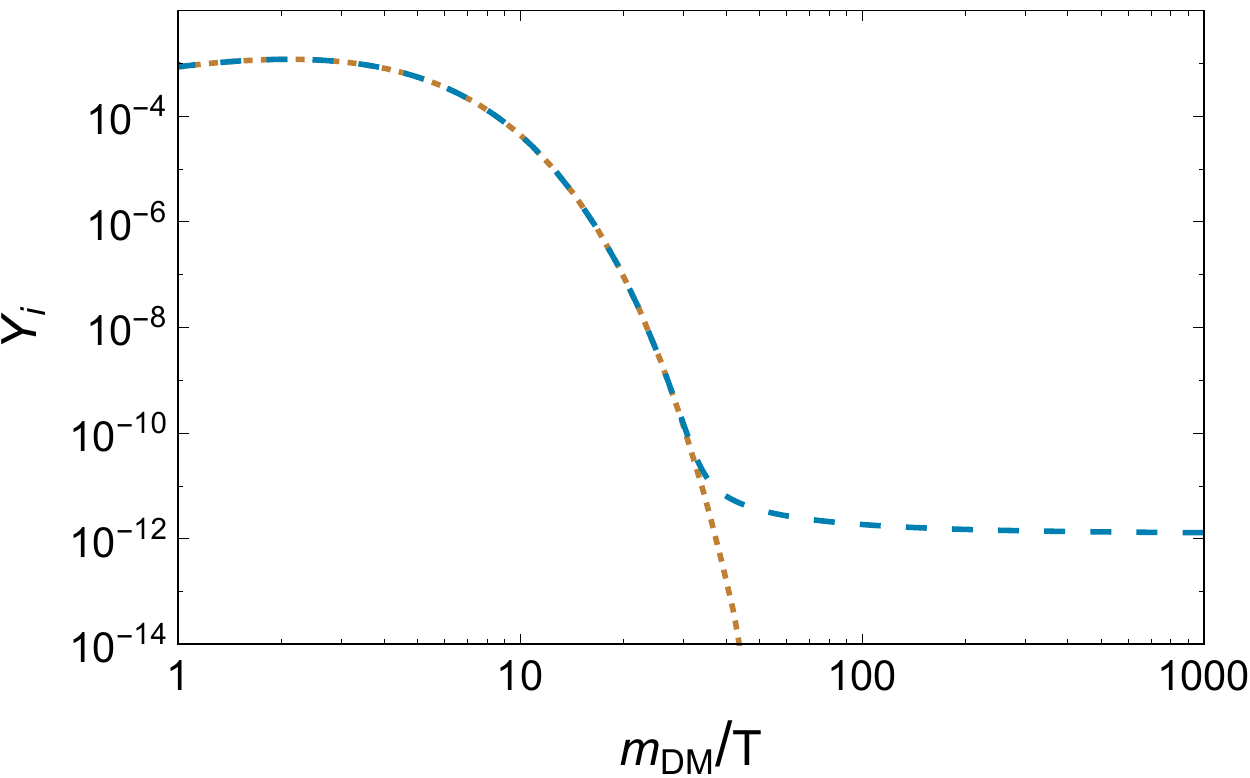}}
		\subfigure[]{\includegraphics[width=0.5\linewidth]{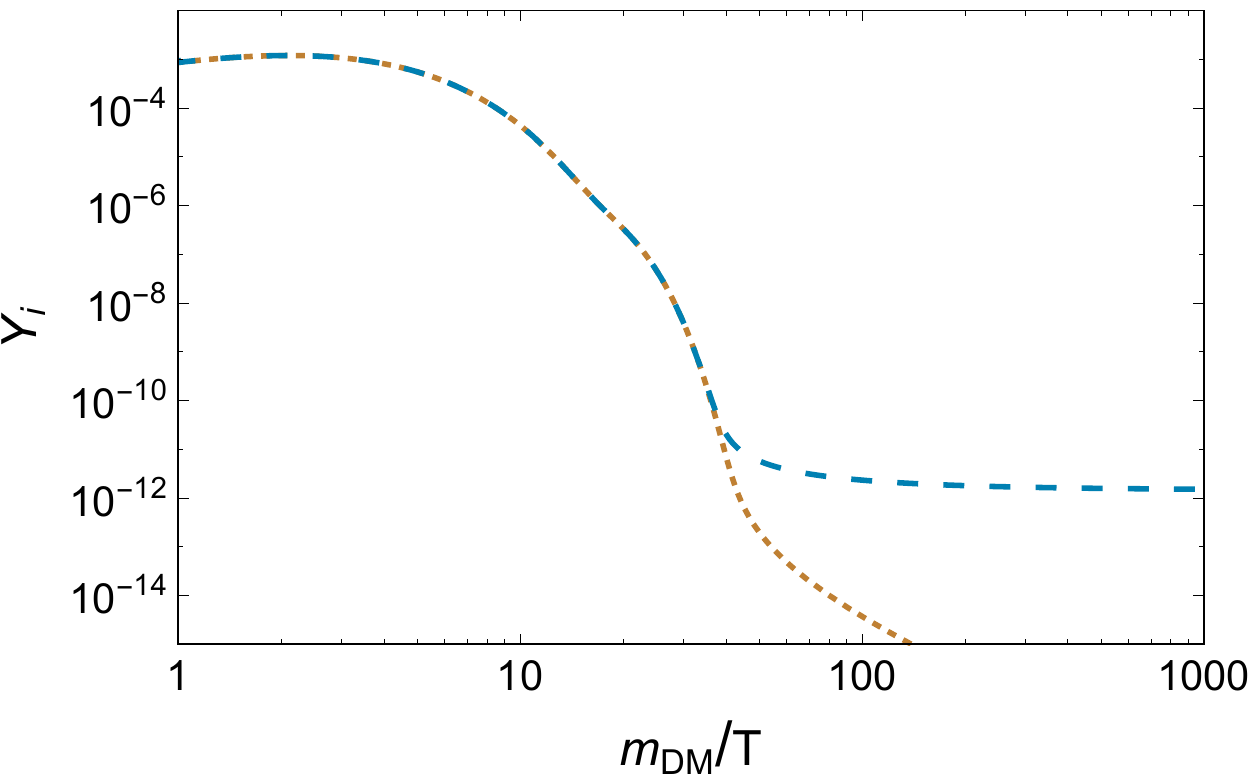}}}
		\mbox{\hskip -20 pt
			\subfigure[]{\includegraphics[width=0.5\linewidth]{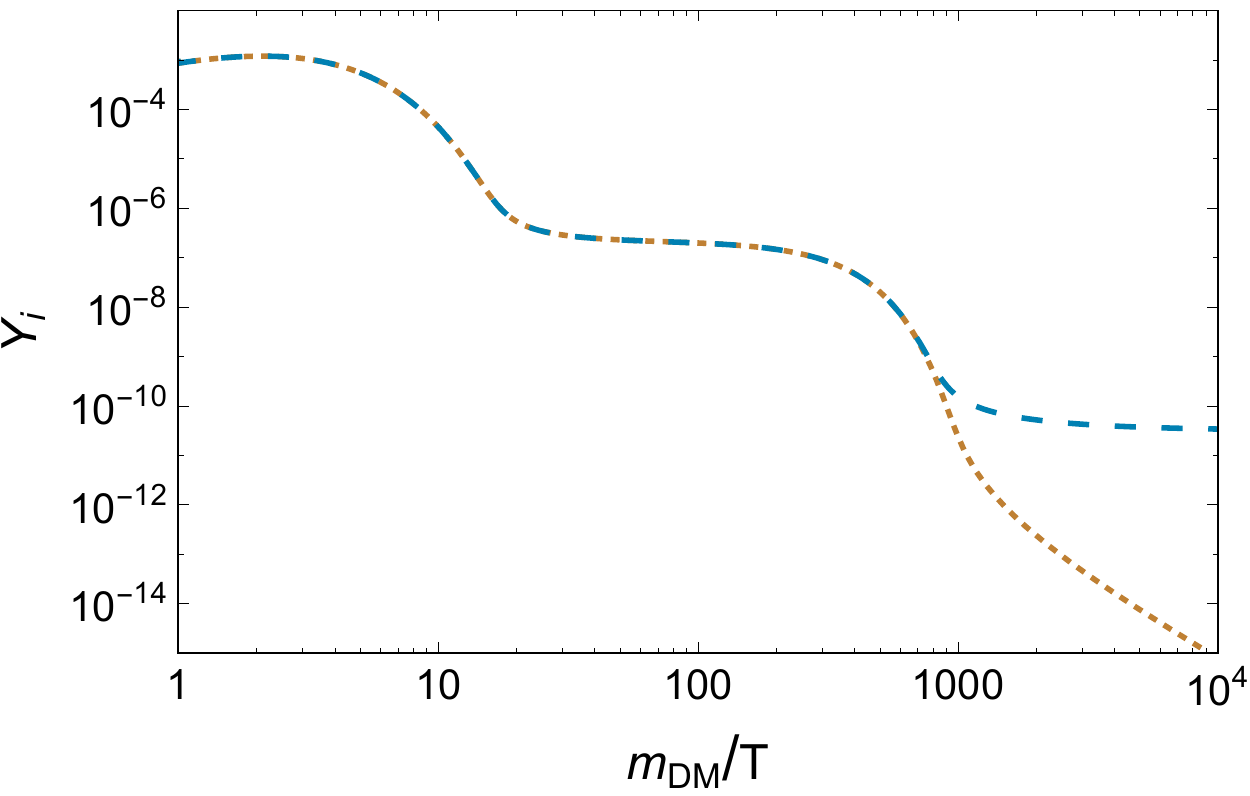}}
			\subfigure[]{\includegraphics[width=0.5\linewidth]{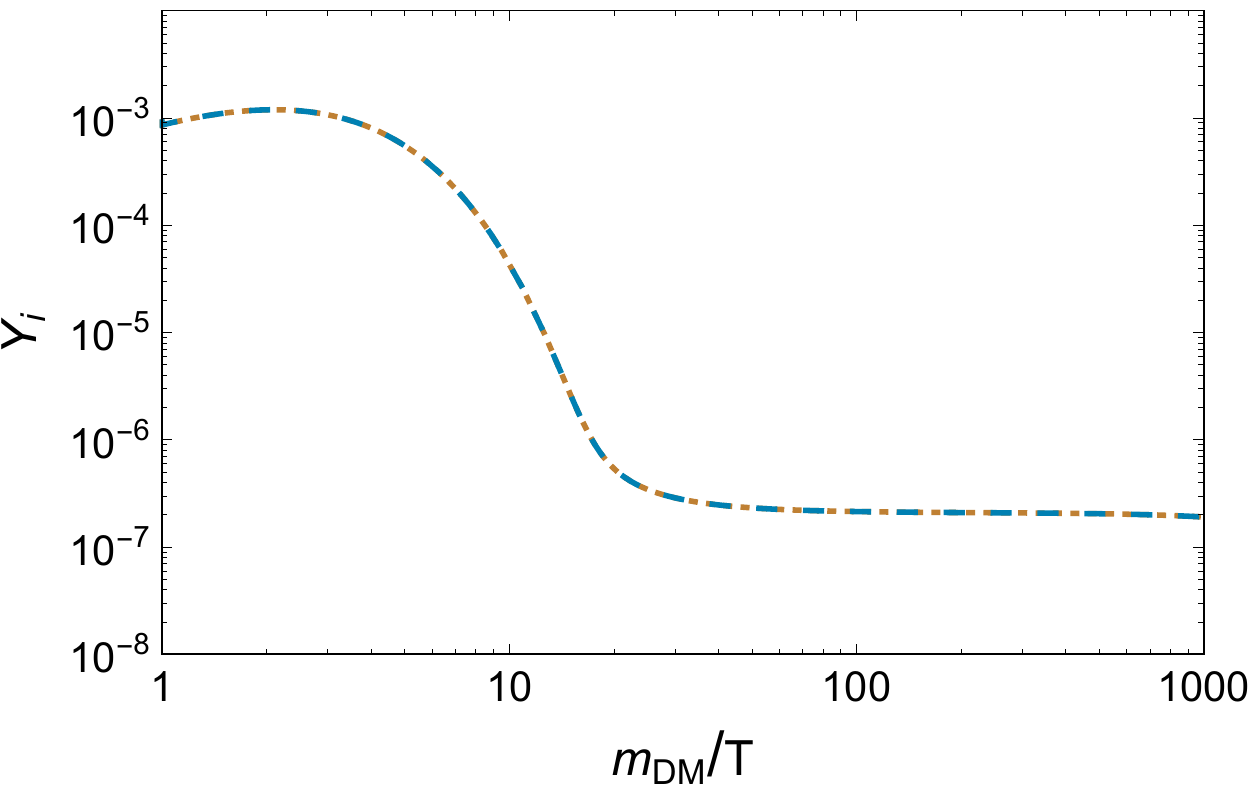}}}
		\caption{The actual number of $\phid$ and $N$ per comoving volume are shown in blue dashed and brown dotted curves, respectively. The panels (a)--(d) are obtained by solving the coupled Boltzmann equations (Eqs.~\eqref{eq:dYDM} and \eqref{eq:dYN}) with the total decay width $\Gamma$ of $N$ as $10^{-10} \gev$, $10^{-15}\gev$, $10^{-18}\gev$ and $0 \gev$, respectively. The effect of decay term is prominent from the plots. The other parameters are chosen as follows: $m_{DM}=140\gev$, $m_{N}=100\gev$, $m_{S_0}=300\gev$, $m_{h}=125\gev$ and $m_{Z^\prime}=3\tev$.}
		\label{fig:1}
	\end{center}
\end{figure}

To avoid the overproduction of the DM, we consider the effect of the decay and inverse decay term of right-handed neutrino in the Boltzmann equations. The decay of $N$ to the SM particle is governed by the Yukawa coupling $y_N$. We vary the coupling $y_N$ for different values of the total decay width $\Gamma$ of the right-handed neutrino, by choosing all other masses as $m_{DM}=140\gev$, $m_{N}=100\gev$, $m_{S_0}=300\gev$, $m_{h}=125\gev$ and $m_{Z^\prime}=3\tev$. 
The results due to the variation of $\Gamma$ for the values of $10^{-10} \gev$, $10^{-15}\gev$, $10^{-18}\gev$ and $0 \gev$ are shown in Fig.~\ref{fig:1}.
The blue dashed and brown dotted curves represent the actual number of $\phid$ and $N$ per comoving volume, respectively. For larger values of the Yukawa coupling $y_N$ i.e., larger decay width, the decay term of $N$ in Eq.~\eqref{eq:dYN} dominates over the other interactions of $N$ before the annihilation effect of $N$ becomes weaker than the dilution effect of the Universe expansion. Hence due to the decay effect, $N$ remains in thermal bath for much longer time compared to the case when there is no decay term present in the analysis. Thus for this case, $N$ can remain in thermal bath continuously before the DM candidate \phid decouples and the result can be seen from Figs.~\ref{fig:1}(a) and \ref{fig:1}(b), which reproduce the standard result assuming $N$ in thermal equilibrium. In other words, for this parameter space, the combination of the interaction of $N$ and its decay effect gives back the result obtained by solving single Boltzmann equation where $N$ assumes to be in equilibrium. On the other hand, as decay width decreases, the decay effect is negligible in the early stage when both $N$ and $\phi_{\rm DM}$ decouple first from the annihilation effect and then the DM relic density is depleted further by the decay effect coming in later as in Fig.~\ref{fig:1}(c). For much smaller decay rate, its effect never becomes effective leaving the DM relic density as in Fig.~\ref{fig:1}(d) with $\Gamma=0$. 
In Ref.~\cite{Fujii:2001xp} a similar qualitative behavior of $ \Yd$ solution is found in a completely different context.

The relic abundance of the DM candidate \phid can be evaluated by,
\begin{align}
	\label{eq:relic}
	\Omega h^2 = \frac{m_{DM} s_0 \Yd(\infty)}{\rho_c/h^2},
\end{align}
where $s_0=2890$ cm$^{-3}$ is the current entropy density of the Universe and $\rho_c/h^2=1.05\times 10^{-5} \gev/ $cm$^3$ is the critical density. $\Yd(\infty)$ is the asymptotic value of the actual number of \phid per comoving volume obtained from numerical solutions of the corresponding Boltzmann equations.

We calculate the velocity-averaged DM annihilation cross-section for the contribution to $NN$ final state through $s$-channel $S_0$ exchange (Fig.~\ref{dia:relic}(a)). The leading term i.e., the $s$-wave contribution in the non-relativistic limit $s= 4 m_{DM}^2$ is
\begin{align}
	\langle \sigma v \rangle_{\phid\phidst\to NN} \equiv \langle \sigma v \rangle_0= \frac{ \lambda_N^2 \lambda_{DS}^2 v^{\prime 2}}{64 \pi m_{DM}^2} \sqrt{1-\frac{m_N^2}{m_{DM}^2}} \frac{m_{DM}^2 - m_N^2}{\left(4m_{DM}^2 - m_{S_0}^2\right)^2 +m_{S_0}^2 \Gamma_{S_0}^2 },
\end{align}
where $\Gamma_{S_0}$ is the total decay width of the scalar $S_0$. 
As discussed earlier, $S_0$ interacts to the right-handed neutrinos dominantly, the total decay width of $S_0$ is saturated by its decay to $N$ pair and/or \phid pair which is given by
\begin{align}
	\Gamma_{S_0}= \frac{ m_N^2 m_{S_0} }{16 \pi v^{\prime 2}} \left(1- \frac{4 m_N^2}{m_{S _0}^2}\right)^{3/2}\text{cos}^2\alpha \, + \frac{ \lambda_{D S}^2 {v^\prime}^2 }{64 \pi m_{S_0}} \left(1- \frac{4 m_{DM}^2}{m_{S _0}^2}\right)^{1/2}.
\end{align}

\begin{figure}[!t]
	\begin{center}
		\mbox{\hskip -20 pt\subfigure[]{\includegraphics[width=0.5\linewidth]{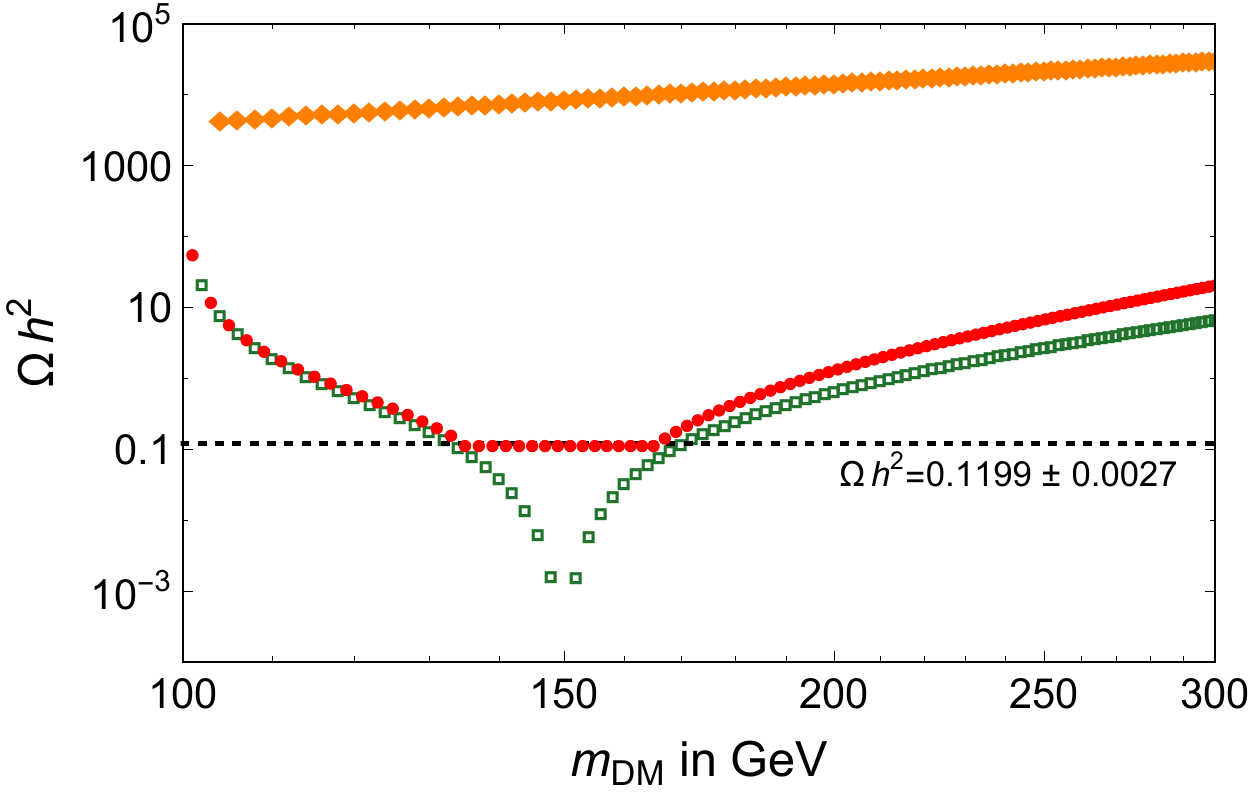}}
			\subfigure[]{\includegraphics[width=0.48\linewidth]{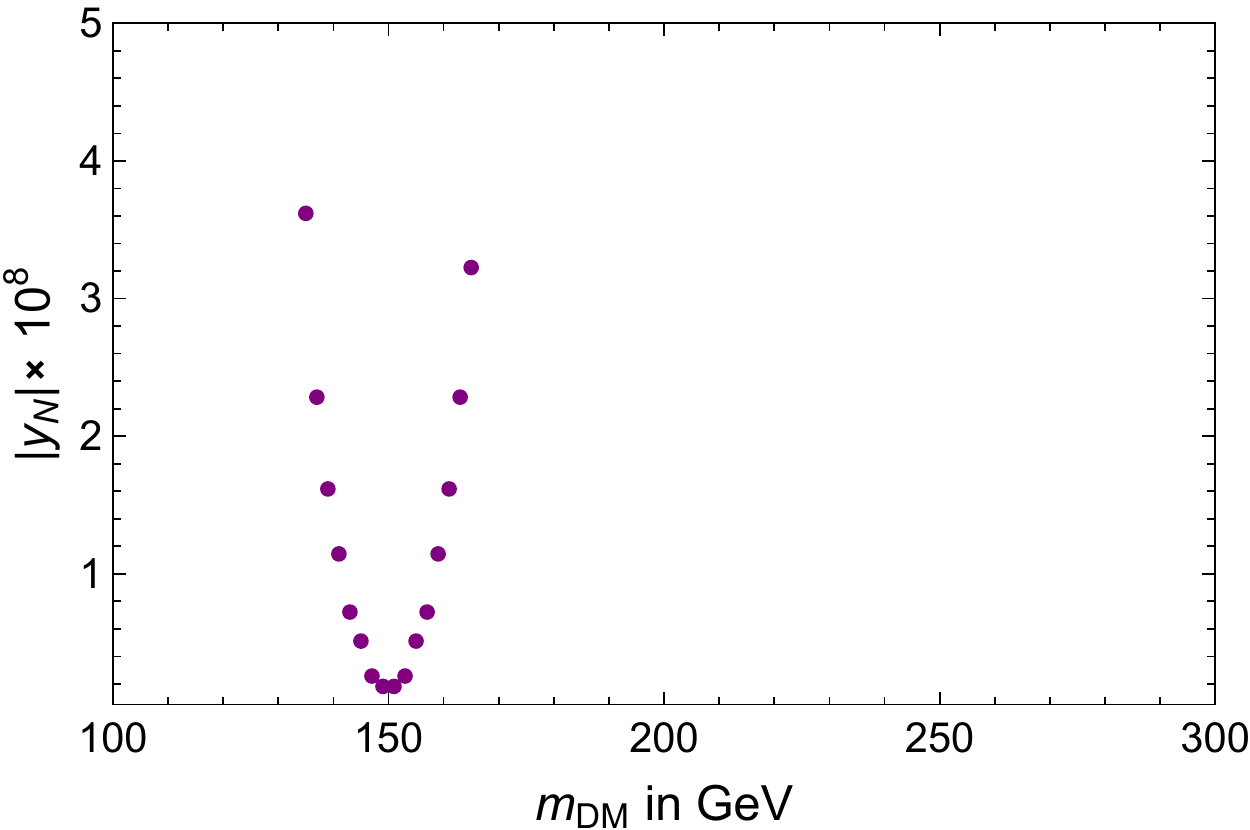}}}
		\caption{(a) The comparison of relic density obtained for three different scenarios is shown as a variation of the DM mass $m_{DM}$. The red points denote the proper solutions of the coupled Boltzmann equations \eqref{eq:dYDM} and \eqref{eq:dYN}. The orange points represent the solutions assuming the vanishing decay rate and the green squares correspond to the solutions for the right-handed neutrino $N$ in thermal equilibrium.  By adjusting the Yukawa coupling $y_N$, the observed relic density can be satisfied over a $30\,$\gev~  range around resonance region of the $B-L$ scalar $S_0$. The masses for the other parameters are chosen as $m_{N}=100\gev$, $m_{S_0}=300\gev$, $m_{h}=125\gev$, $m_{Z^\prime}=3$TeV and $\lambda_{DS}=0.3$. (b) The variation of $y_N$ with $m_{DM}$ corresponding to the red points, which satisfies observed relic density, in the left panel figure where the $N$ decay effect is included in the Boltzmann equations. A larger coupling $y_N$ i.e., larger decay width is needed to satisfy relic away from the $S_0$ resonance region. }
		\label{fig:reso}
	\end{center}
\end{figure}

To illustrate the effects of three different scenarios, we show the variation of thermal relic density with the DM mass $m_{DM}$ in Fig.~\ref{fig:reso}(a), by choosing a benchmark point 
%
$m_{N}=100\gev$, $m_{S_0}=300\gev$, $m_{h}=125\gev$, $m_{Z^\prime}=3\tev$ and $\lambda_{DS}=0.3$. We mention that Fig.~\ref{fig:reso} is highlighted for a particular benchmark point and it is easily understood that the observations made here can be realized in other regions in the parameter space as well. The green squares correspond to the case where the right neutrino $N$ is assumed to be in thermal equilibrium and hence the result is obtained by solving only one Boltzmann equation. It can be seen that for most of the parameter space the annihilation rate is small and can not explain the observed relic abundance by Planck data $\Omega h^2 = 0.1199 \pm 0.0027$ ~\cite{Ade:2015xua} (shown in the black dotted line). The criteria for correct relic abundance is satisfied only at the near resonance region of the scalar $S_0$, virtually at two points one before and the one after the $S_0$ resonance. The orange solid squares depict the case when interactions of $N$ are considered in the theory. The two coupled Boltzmann equations (Eqs.~\eqref{eq:dYDM} and \eqref{eq:dYN}) are solved without the decay effect of $N$. As $N N\to Z^\prime \to  f \bar{f}$ annihilation rate is suppressed due to the large $Z^\prime$ mass, $N$ decouples earlier than the DM and hence this scenario corresponds to overabundance of the DM particles. The situation improves significantly after incorporating the decay effect of $N$ which can be seen from the red points. By adjusting the Yukawa coupling $y_N$, we can satisfy the relic density over a $30\,$\gev  range around resonance region of the $B-L$ scalar $S_0$. Figure~\ref{fig:reso}(b) illustrates the variation of $y_N$ with $m_{DM}$ corresponding to the red points, which satisfy observed relic density, in the left panel figure. A larger coupling $y_N$ i.e., larger decay width is needed to satisfy relic away from the $S_0$ resonance region.

It can be seen that from the curve with green squares, there exist two positions, one before and the one after the $S_0$ resonance, where the observed relic abundance is satisfied. At the top of the resonance due to the huge enhancement of $\phid \phidst\to S_0 \to NN$ annihilation cross-section the relic density is very suppressed $\sim {10}^{-5}$. Similar observation was made in Ref.~\cite{Rodejohann:2015lca} and as discussed that it is difficult to find viable models right on top of the resonance due to the suppression of relic density, we show that by incorporating the decay effect of right-handed neutrino $N$ in the analysis, one can evade such problem.

\section{Constraints from direct detection experiments}
\label{sec:directD}

\begin{figure}[!t]
	\begin{center}
		\mbox{\hskip -20 pt\subfigure[]{\includegraphics[width=0.5\linewidth]{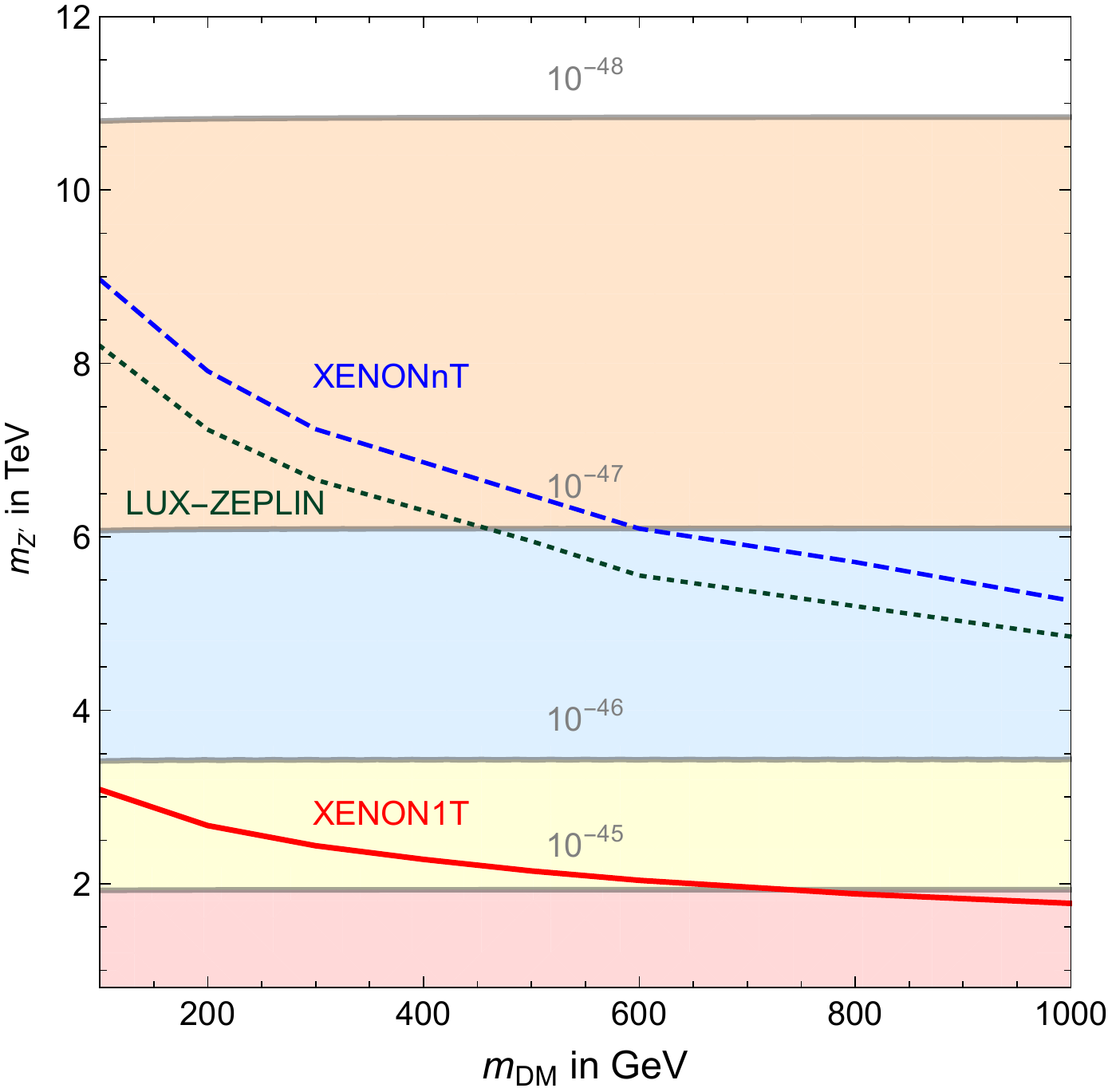}}
	\subfigure[]{\includegraphics[width=0.5\linewidth]{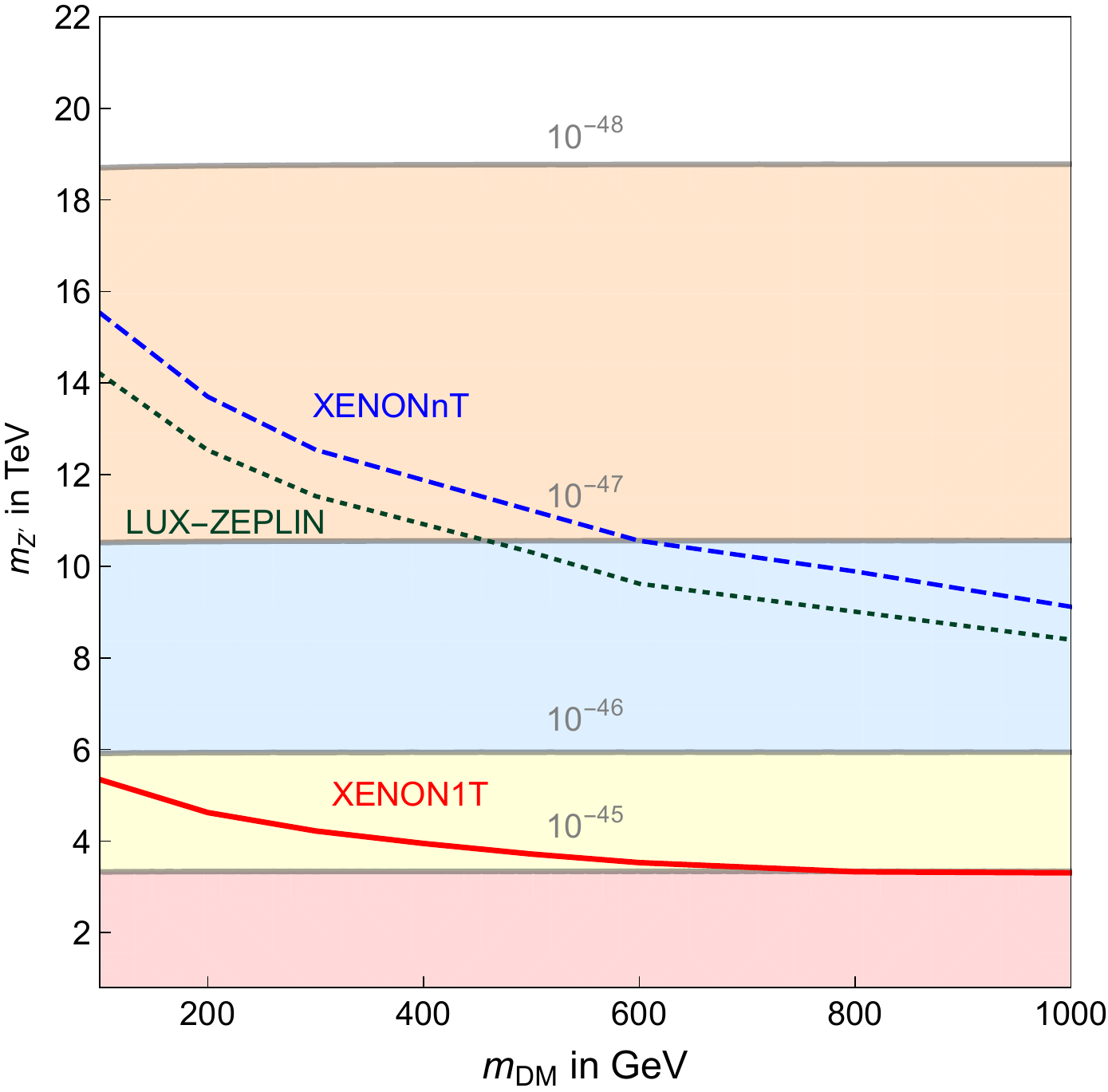}}}
\caption{The contour plot for direct detection cross-section through a $t$- channel $Z^\prime$ exchange is shown in $m_{DM}- m_{Z^\prime}$ plane. The left panel (a) corresponds to $\qd\!=\!1/2$ and the right panel (b) is obtained for $\qd\!=\!3/2$. The red, yellow, blue and orange shaded regions denote cross-section greater than $ 10^{-45}$ cm$^2$, within $ 10^{-45}$cm$^2-10^{-46}$ cm$^2$, within $ 10^{-46}$cm$^2- 10^{-47}$cm$^2$ and within $ 10^{-47}$cm$^2- 10^{-48}$cm$^2$, respectively. The red solid, green dot and blue dashed curves are the current or future bounds obtained in XENON1T~\cite{Aprile:2017iyp}, LUX-ZEPLIN~\cite{Mount:2017qzi} and XENONnT~\cite{Aprile:2015uzo} experiments, respectively. The region below the mentioned curves are excluded at 90\% confidence level. We assume the $B-L$ gauge coupling $g_{BL}=0.3$ for both the panels.}
		\label{fig:Direct}
	\end{center}
\end{figure}

In this section we discuss the bound from the DM direct detection experiments. In the model under consideration, the DM is scalar particle and interacts with the nucleons either via $t$- channel exchange of the gauge boson $Z^\prime$ or the SM Higgs boson $h$. The contribution from $h$ exchange to the scattering cross-section depends strongly on the DM-Higgs coupling $\lambda_{D H}$. The value of $\lambda_{D H}$ is not of relevance for the purpose of our paper. Hence in this section we restrict ourselves on the DM-nuclei interaction through $Z^\prime$ only.

The effective Lagrangian describing the scattering off the scalar DM particle \phid with the nucleon by $Z^\prime$ mediated channel is
\begin{align}
\mathcal{L}_\text{eff}= i\frac{\qd g_{BL}^2}{3 m_{Z^\prime}^2} V^\mu \bar{q} \gamma_\mu q
,\qquad q\in \{u,d\},
\end{align}
where $V^\mu$ is the vector current arising from the kinetic term of the DM particle \phid. Decomposing \phid in terms of real and imaginary components as \phid$= \left(\phi_1+ i\phi_2\right)/\sqrt{2}$, we get $V^\mu\simeq \left(\phi_1 \partial^\mu \phi_2- \phi_2 \partial^\mu \phi_1 \right)$.

The SI DM-nuclei scattering cross-section for the scalar DM mediated by a $t$-channel gauge boson $Z^\prime$ is~\cite{Goodman:1984dc} 
\begin{align}
\sigma_{SI}^N= \frac{1}{16 \pi} \left(\frac{M_N \,m_{DM}}{M_N+ m_{DM}} \right)^2 |b_N|^2,
\end{align}
where $M_N$ is the mass of the nuclei and the coefficient $b_N$ is given by
\begin{eqnarray}
b_N= \left(A-Z\right) b_n + Z b_p,~~~b_n= b_u +2 b_d,~~~b_p= 2b_u + b_d,
\end{eqnarray}
with $Z$ and $A$ are the atomic and mass number of the nuclei, respectively.
In terms of our model parameters $$ b_n=b_p= i\frac{\qd g_{BL}^2}{ m_{Z^\prime}^2}.$$
Thus the SI scattering contribution for the DM and a single nucleon with mass $M_n$ is
\begin{align}
\label{eq:Dcross}
\sigma_{SI}^{\tiny{Z^\prime}}= \frac{1}{16 \pi} \left(\frac{M_n \,m_{DM}}{M_n+ m_{DM}} \right)^2 \frac{\qd^2 g_{BL}^4}{ m_{Z^\prime}^4}.
\end{align}

In Fig.~\ref{fig:Direct}, the predictions for SI scattering cross-section $\sigma_{SI}^{\tiny{Z^\prime}}$ is shown  in the plane $m_{DM}- m_{Z^\prime}$. We assume the $B-L$ gauge coupling $g_{BL}=0.3$. The left panel (a) corresponds to $\qd\!=\!1/2$ and right panel (b) is obtained for $\qd\!=\!3/2$. The red, yellow, blue and orange shaded regions denote cross-section greater than $ 10^{-45}$ cm$^2$, within $ 10^{-45}$cm$^2-10^{-46}$ cm$^2$, within $ 10^{-46}$cm$^2- 10^{-47}$cm$^2$ and within $ 10^{-47}$cm$^2- 10^{-48}$cm$^2$, respectively. The red solid, green dot and blue dashed curves are the current or future bounds obtained in XENON1T~\cite{Aprile:2017iyp}, LUX-ZEPLIN~\cite{Mount:2017qzi} and XENONnT~\cite{Aprile:2015uzo} experiments, respectively, where the region below the mentioned curves are excluded at 90\% confidence level. It can be inferred that for the DM mass below $1\tev$, the direct detection limit from XENON1T experiment is strongly competing with the current collider bounds on the mass of the $Z^\prime$ gauge boson. It should be noted from Eq.~\eqref{eq:Dcross} that due to the proportionality of $\sigma_{SI}^{\tiny{Z^\prime}}$ on $\qd^2$ the bound on $m_{Z^\prime}$ is stronger when the $B-L$ charge of \phid increases.

\section{Right-handed neutrino phenomenology}
\label{sec:RHnu}

In the model under consideration, the right-handed neutrinos are introduced to assure the $B-L$ gauge anomaly cancellation. The small neutrino mass terms are generated via Type-I seesaw mechanism as can be seen from Eq.~\eqref{lag}. The right-handed neutrinos are SM gauge singlet but are charged under $B-L$ as shown in Table~\ref{Table}. 
As discussed in Sec.~\ref{sec:model}, only one of the three $N_i$'s is lighter than the DM candidate, in this section we consider the phenomenology of the mentioned right-handed neutrino $N$ with a mass $\le \mathcal{O}$(TeV).
The coupling $y_N$ of the right-handed neutrino $N$ with the SM leptons are Yukawa type, which governs its decay in three possible channels and the decay widths are given in Eqs.~\eqref{eq:Ndecayh}-\eqref{eq:NdecayZ}. However, the production of right-handed neutrino $N$ is dictated by its $B-L$ charge and the gauge coupling, which is electro-weak in nature. Below we discuss the production and decays of the right-handed neutrino at the LHC.

\begin{figure}[h]
	\begin{center}
		\includegraphics[width=0.5\linewidth]{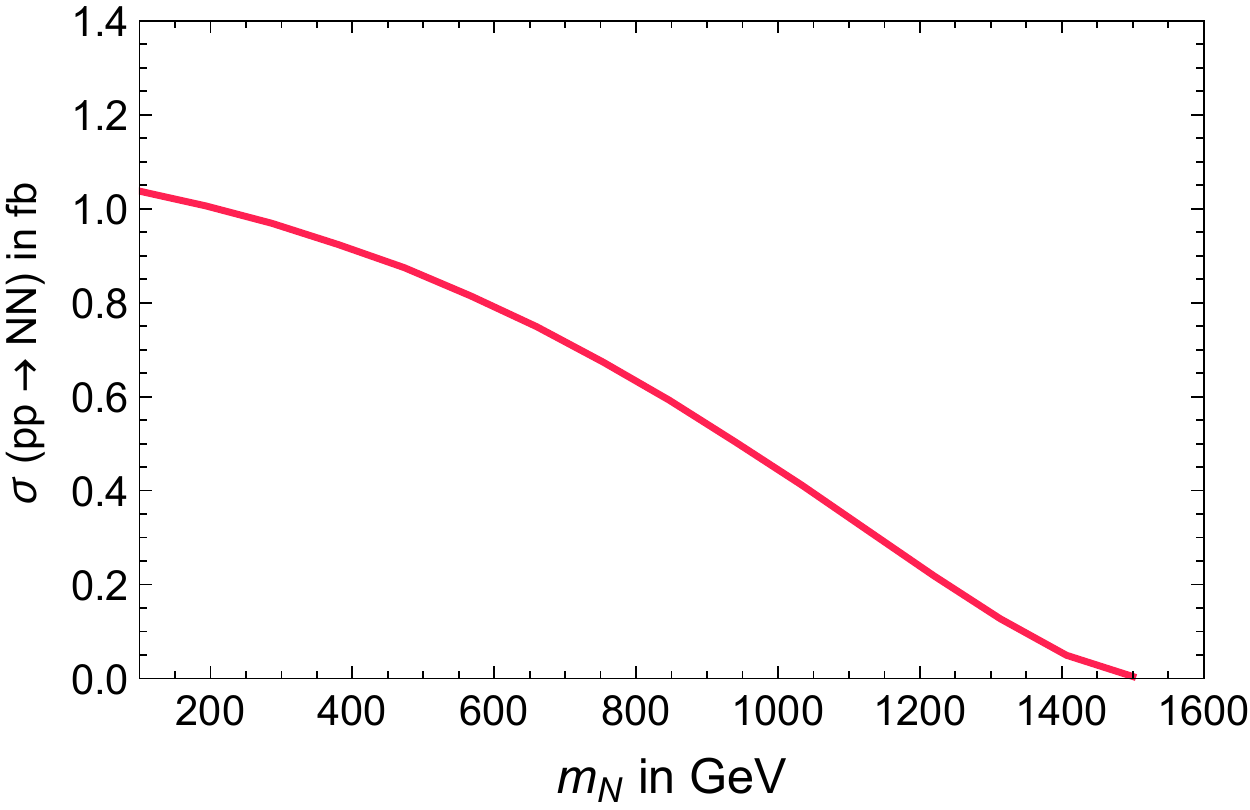}
		\caption{The variation of production cross-section of right-handed neutrino $N$ pair with its mass $m_N$, at 14 \tev LHC. The other parameters are chosen such that $m_h= 125 \gev$, $m_{S_0}=354 \gev$, $m_{DM}=178 \gev$, $m_{Z^\prime}= 3 \tev\!$ and $g_{BL}=0.3$.}
		\label{fig:sigmaN}
	\end{center}
\end{figure}
\subsection{Production}
The dominant production mode for $N$ pair is in the $s$-channel via $Z^\prime$ gauge boson. Due to the mass bound of $Z^\prime$, i.e., $m_{Z^\prime}\geq 2.8$ TeV \cite{z'bnd} from LHC, such mode is suppressed. Nevertheless, it is the only dominant mode available for the production. 
In Fig.~\ref{fig:sigmaN} we present the pair production cross-section of right-handed neutrino $N$ at the LHC with 14 TeV $E_{CM}$ with a choice of PDF as CTEQ6L \cite{cteq}. The renormalization and factorization scale is chosen to be $\sqrt{\hat{s}}$. The other parameters are fixed at $m_h= 125 \gev$, $m_{S_0}=354 \gev$, $m_{DM}=178 \gev$, $m_{Z^\prime}= 3 \tev$ and $g_{BL}=0.3$. Since the coupling is electro-weak in nature and due to heavy $Z^\prime$ exchange, we can see that the cross-section merely turns out be $\mathcal{O}$(fb). 

\subsection{Decay}
\begin{figure}[h]
	\begin{center}
		\includegraphics[width=0.48\linewidth]{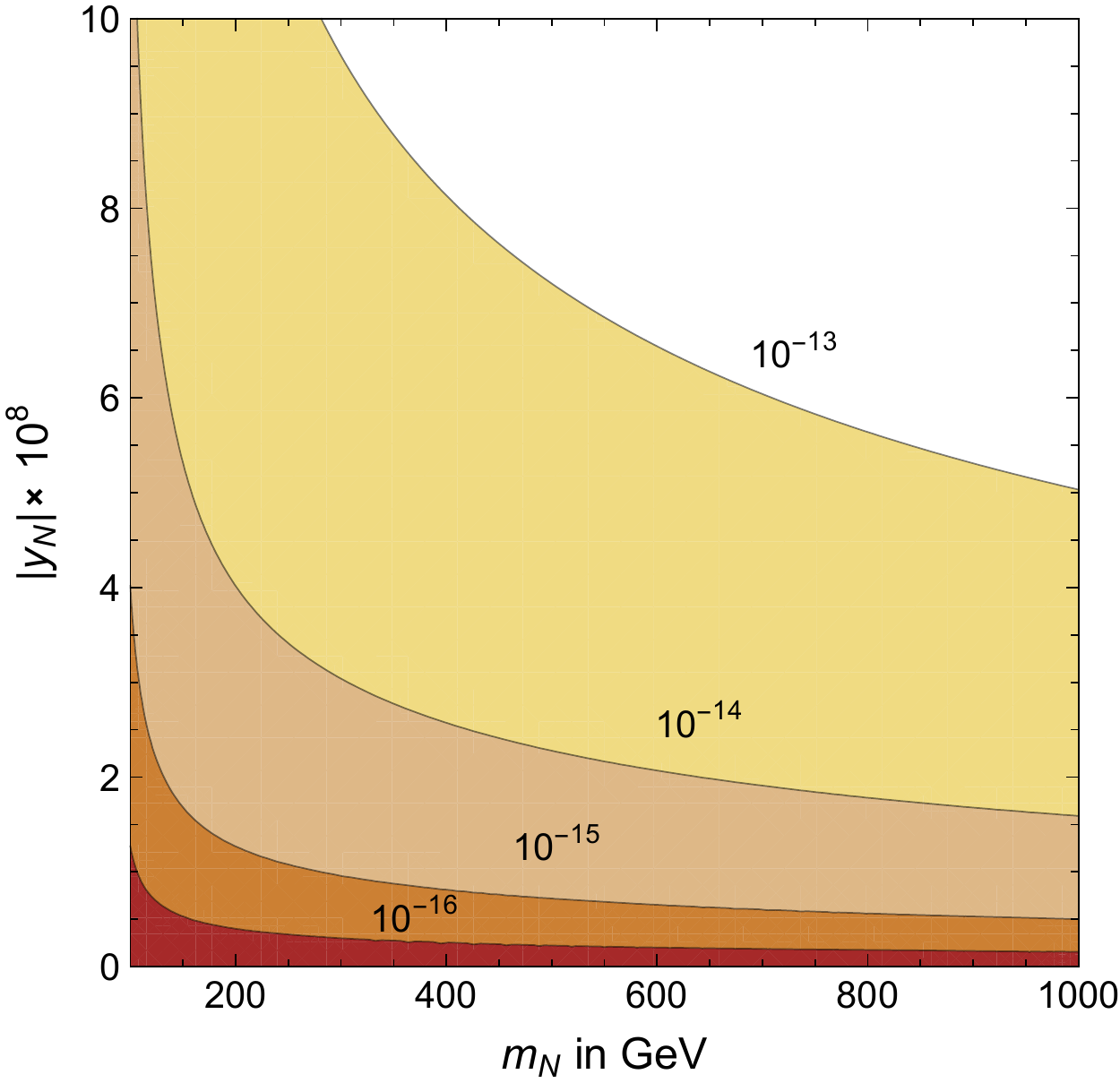}
		\caption{The total decay width of the right-handed neutrino is shown as a variation in $m_N - y_N$ plane. The different regions of the decay width from $10^{-16}$ GeV to $10^{-13}$ GeV are shown in red, orange, brown and yellow, respectively.}
		\label{fig:totdcdth}
	\end{center}
\end{figure}
The right-handed neutrino can decay into the modes with SM gauge bosons and leptons via mixing which is proportional to the Yukawa coupling $y_N^2$, and the decay widths for the mentioned modes can be found in Eq.~\eqref{eq:Ndecayh}-\eqref{eq:NdecayZ}. We have seen in Sec.~\ref{sec:relic} that, to satisfy the observed relic density a tiny value of $y_N \sim 10^{-8}$ is needed for a right-handed neutrino mass $m_N\sim \mathcal{O}(100)$GeV.
Such a low Yukawa coupling slows the decay rate which in turn gives rise to displaced decays of the right-handed neutrino.  In Fig.~\ref{fig:totdcdth} we show the total decay width of the right-handed neutrino as a variation of its mass $m_N$ and Yukawa coupling  $y_N$. The different shaded regions of the decay width from $10^{-16}$ GeV to $10^{-13}$ GeV are shown in red, orange, brown and yellow, respectively. It is evident from Fig.~\ref{fig:totdcdth} that there is significant region of parameter space that can be explored in the collider searches which comprise of displaced vertex signatures. Decays of such right-handed neutrino into charged leptons and gauge bosons leave displaced charged track at the collider. In the following collider study we search for such displaced final states at the LHC with 14 TeV $E_{CM}$ by choosing some suitable benchmark points.

\subsection{Benchmark points and collider signature}
We choose two benchmark points defined in Table~\ref{Table:2} to investigate the collider phenomenology of the right-handed neutrino $N$ at the LHC. The benchmark points are chosen such a way that they satisfy the observed relic density by the DM annihilation via $s$-channel $S_0$ exchange and also have displaced decays for the right-handed neutrino $N$. The BP1 deals with relatively lighter DM particle and right-handed neutrino masses $176\gev$ and $110\,$GeV, compared to BP2, where the masses are $600\gev$ and $500\,$GeV, respectively.  Below we explore the effect of different mass spectrum in the kinematics of the decay products of the right-handed neutrinos $N$. 
\begin{table}[h]
	\centering
	\begin{tabular}{ |c| c |c |c|c|c|}
		\hline \hline
		& $m_{h}$ & $m_{S_0}$ &  $m_{DM}$ & $m_N$ & $m_{Z^\prime}$ \\ \hline
		BP1 & $125 \gev$ & $ 300 \gev$ & $165 \gev $ & $110 \gev$ & $3 \tev$  \\
		\hline
		BP2 & $125 \gev$ & $ 1225 \gev$ & $600 \gev $ & $500 \gev$ & $3 \tev$  \\
		\hline
		\hline 
	\end{tabular}
	\caption{Masses of different particles for two benchmark points.}  \label{Table:2}
\end{table}

\begin{table}[b]
	\begin{center}
		\renewcommand{\arraystretch}{1.4}
		\begin{tabular}{||c|c|c||}
			\hline\hline
			Branching &BP1&BP2\\
			
			fractions of $N$&&\\
			\hline
			$W^\pm e^\mp$& 79\%& 51.6\%\\
			\hline
			$Z \nu$ & 21\%& 25.7\%\\
			\hline
			$h\nu $&--& 22.7\%\\
			\hline
			\hline
		\end{tabular}
		\caption{\label{NBr} Branching fractions of the right-handed neutrino $N$ to different decay modes for BP1 and BP2 where the total decay widths are $1.09\times {10}^{-15}$\gev and $1.74\times {10}^{-14}$\gev, respectively.}
	\end{center}
\end{table}

The right-handed neutrino with lighter mass (in BP1) decays mainly to gauge boson modes, i.e., $Z\nu$ and $W^\pm e^\mp$. For higher mass of $m_N$ (in BP2), $N$ decaying to $h\nu$ and $S_0\, \nu$ are also feasible. The decay branching fractions are given in Table~\ref{NBr} for the two benchmark points and it is visible that $h\nu$ mode is open only for  BP2, where all three modes share the branching fractions almost equally. However, for BP1 $W^\pm e^\mp$ is the most dominating mode. The choice of mass spectra in the benchmark points didn't allow its decay to $S_0\, \nu$  via mixing. The decay widths of $N$ are proportional to $y^2_N$, as can be seen from Eqs.~\eqref{eq:Ndecayh}-\eqref{eq:NdecayZ}. Later we discuss that such small couplings can cause displaced decays of $N$, which give rise to displaced charged leptons or bosons. Looking into the  decay of  $N \to W^\pm e^\mp$, it can easily be understood that the $W^\pm$ will go through prompt decays which will give rise to either two jets or one charged lepton. Thus displaced $2\ell,~ 3\ell~ \rm{and} ~4\ell$ predicted by the mentioned decays are the golden channels to look for at the LHC.

There are other studies in the context of the displaced decays of right-handed neutrinos for Type-I seesaw where the displaced decay width of right-handed neutrino is proportional to the corresponding Yukawa couplings $y_N^2$ \cite{ty1d}.  The situation however changes  a lot in the context of supersymmetry as the superpartners of right-handed neutrino i.e., the right-handed sneutrino can also undergo displaced decays via the mixing with left-handed sneutrinos. In some parameter space, such mixing angles go to very small values caused by cancellation in the parameter space apart from the Type-I seesaw type suppression. This prompts the displaced decays of such right-handed neutrinos into charged and neutral leptons \cite{LFVpheno,ty1s}.

For the model under consideration, we simulate the right-handed neutrino events, pair produced at  the LHC with displaced charged leptons final states. We used CalcHEP, PYTHIA \cite{calchep,pythia}  for the event generation and simulation. The jet formation has been performed using the {\tt Fastjet-3.0.3} \cite{fastjet} with the {\tt CAMBRIDGE AACHEN} algorithm. We have selected a jet size $R=0.5$ for the jet formation, with the following criteria:
\begin{itemize}
	\item the calorimeter coverage is $\rm |\eta| < 4.5$
	
	\item the minimum transverse momentum of the jet $ p_{T,min}^{jet} = 10$ GeV and jets are ordered in $p_{T}$
	\item leptons ($\rm \ell=e,~\mu$) are selected with
	$p_T \ge 20$ GeV and $\rm |\eta| \le 2.5$
	\item no jet should be accompanied by a hard lepton in the event
	\item $\Delta R_{\ell j}\geq 0.4$ and $\Delta R_{\ell \ell}\geq 0.2$
	\item Since an efficient identification of the leptons is crucial for our study, we additionally require  
	a hadronic activity within a cone of $\Delta R = 0.3$ between two isolated leptons to be $\leq 0.15\, p^{\ell}_T$ GeV, with 
	$p^{\ell}_T$ the transverse momentum of the lepton, in the specified cone.
\end{itemize}

\begin{figure}
	\begin{center}
		\includegraphics[width=0.38\linewidth,angle=-90]{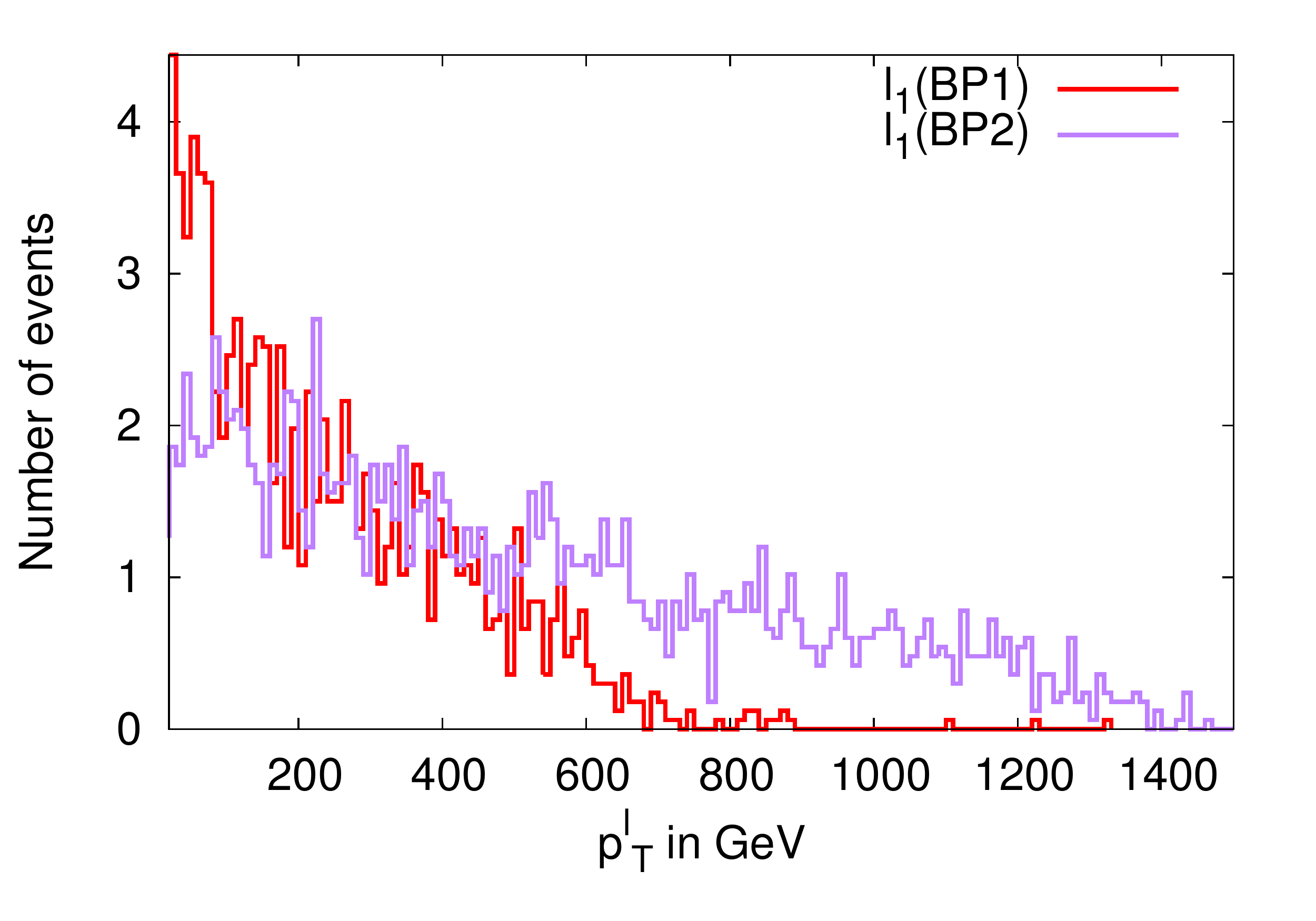}
		\caption{The transverse momentum of the charged leptons arising from the decay of $N$ and the corresponding $W^\pm$ for the two benchmark points at the 14 \tev LHC. The BP1 and BP2 are defined in Table~\ref{Table:2}.}
		\label{fig:lpt}
	\end{center}
\end{figure}

Figure~\ref{fig:lpt} shows the transverse momentum of the charged leptons arising from the decays of $N$ and the corresponding $W^\pm$ for the two benchmark points. The more phase space in the case of BP2 allows the charged lepton to be of very high energy $\mathcal{O}$(TeV), much higher than the BP1 case, which is around few hundreds of GeV. Two such different scenarios can thus be distinguished by applying appropriate lepton $p_T$ cuts and if exist, will be discovered at the LHC with 14 TeV $E_{CM}$. 
\begin{figure}
	\begin{center}
		\mbox{\hskip -20 pt\subfigure[]{\includegraphics[width=0.35\linewidth,angle=-90]{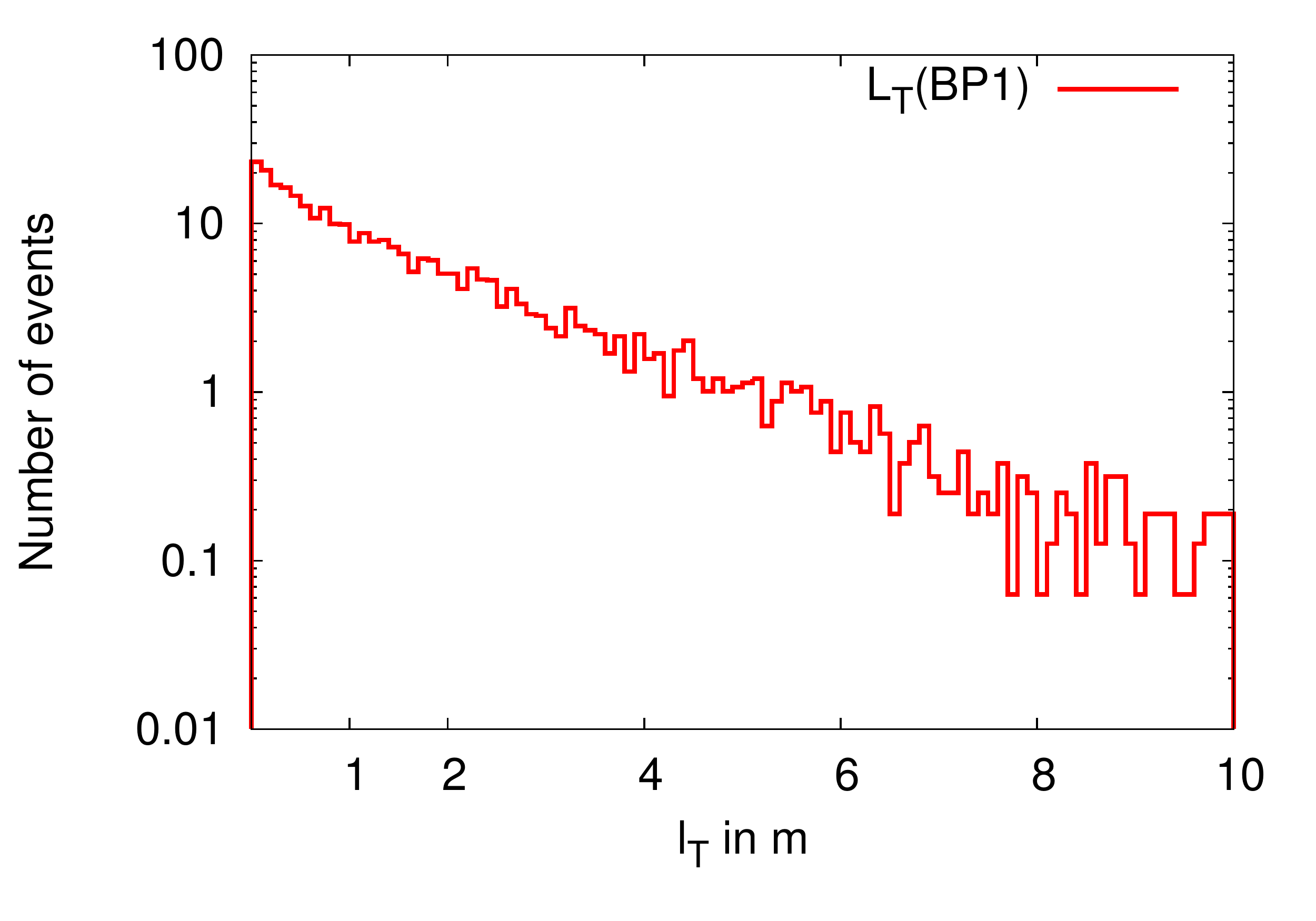}}
		\subfigure[]{\includegraphics[width=0.35\linewidth,angle=-90]{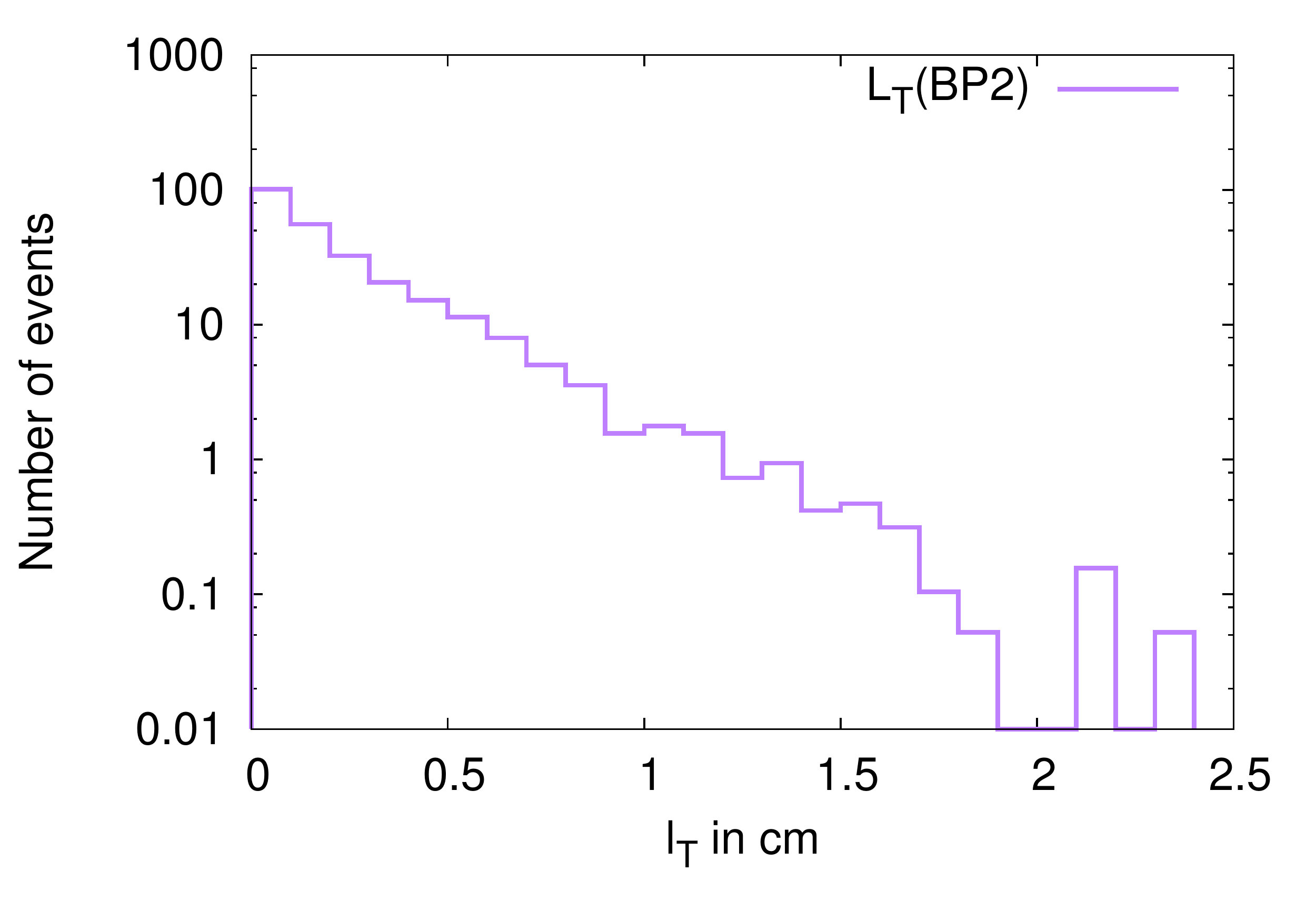}}}
		\caption{The transverse decay length of the right-handed neutrino $N$ in meter for BP1 and in centimeter for BP2 at the 14 \tev LHC, after imposing the basic cuts in a PYTHIA based simulation. The BP1 and BP2 are defined in Table~\ref{Table:2}.}
		\label{fig:histN}
	\end{center}
\end{figure}

The charged leptons arising from the decays of right-handed neutrino $N$ are produced after some travel time of 
$N$, giving rise to displaced charged tracks.  The displaced $W^\pm$s' produced from such decays go through a prompt decay to either charged leptons or quarks, leaving the possibility of another displaced charged track. Figure~\ref{fig:histN}(a) and  Fig.~\ref{fig:histN}(b) show the transverse decay length of the right-handed neutrino $N$ produced at the LHC for BP1 and BP2, respectively. It is evident that the charged track can be seen from few centimeters (for BP2) to few meters (for BP1) length for those particular choice of parameter spaces. Obviously such signals have no SM backgrounds making them completely clean in nature. In Table~\ref{signal} we present the signal numbers at an integrated luminosity of 100 fb$^{-1}$. It can be seen that for $2\ell$ case, we have sufficient events to probe such parameter space. For displaced $3\ell$ and $4\ell$ signals, one has to wait for higher luminosities depending on the choice of benchmark points. Non-observation of such displaced charged tracks clearly put bounds in the $m_{Z'}-m_N$ parameter space at a given luminosity. Below we explore the mentioned parameter space that can be ruled out at the LHC with increasing luminosity.

\begin{table}
	\begin{center}
		\renewcommand{\arraystretch}{1.4}
		\begin{tabular}{||c|c|c||}
			\hline\hline
			Final &BP1&BP2\\
			
			states&&\\
			\hline
			$\geq 2\ell$&33.7&31.6\\
			\hline
			$\geq 3\ell$ &5.5&8.5\\
			\hline
			$\geq 4\ell $&2.9&1.1\\
			\hline
			\hline
		\end{tabular}
		\caption{Final state numbers for $2\ell,~ 3\ell,~ 4\ell$ at the 14 TeV LHC at an integrated luminosity of 100 fb$^{-1}$. }\label{signal}
	\end{center}
\end{table}
\begin{figure*}[!h]
	\begin{center}
	\includegraphics[width=0.5\linewidth]{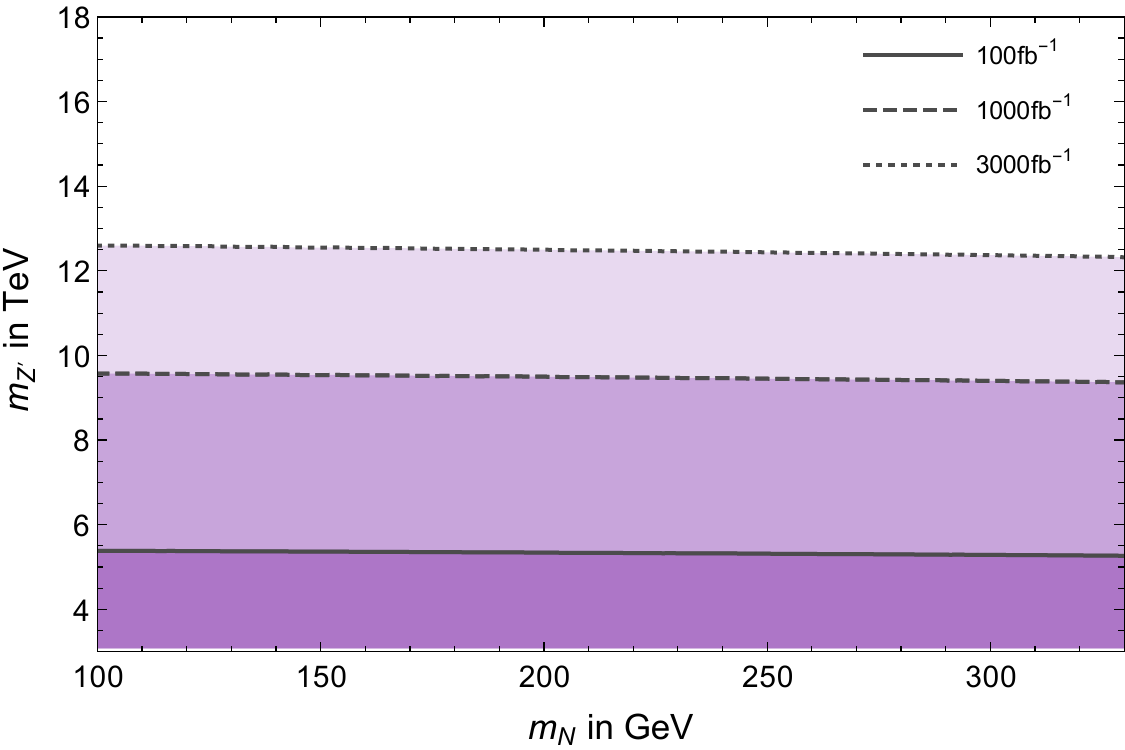}
		\caption{The exclusion limits derived from the non-observation of displaced di-leptonic charge tracks arising from the right-handed neutrino $N$ decays are shown in $m_{N}- m_{Z^\prime}$ plane at 95\% CL for three different luminosities at the $14$ TeV LHC. The deep, lighter and lightest purple regions denote the exclusion regions within the $100\invfb\!, 1000$\invfb  and $3000$\invfb of integrated luminosities, respectively, at the LHC.}
		\label{fig:Luminosity}
	\end{center}
\end{figure*}

Utilizing the non-observation of displaced di-leptonic charge tracks arising from the right-handed neutrino decays, we put bounds on the mass of $Z^\prime$, which is highlighted in Fig.~\ref{fig:Luminosity}.  It depicts the bounds on $m_{Z'}$-$m_N$ plane for non-observation of displaced di-leptons at 95\% CL assuming $30\%$ acceptance as obtained for the mass range of $m_N$ (similar to the case of BP1). At 100 fb$^{-1}$ integrated luminosity, such events exclude $m_{Z'} \lesssim 5.5$ TeV for $m_N \sim 100-400$ GeV.  The region below the solid line (dark purple region) can be ruled out within the integrated luminosities of 100 fb$^{-1}$. Similarly $m_{Z'} \lesssim 9.5$ TeV  can be excluded at 1000 fb$^{-1}$ (dashed line) and $m_{Z'} \lesssim 12.5$ TeV can be ruled out at 3000 fb$^{-1}$  (dotted line) of integrated luminosities at the LHC with 14 TeV $E_{CM}$.

\section{Discussions and conclusions}
\label{sec:conclusion}

In this paper we focus on an extension of the SM where the $B-L$ charged right-handed neutrinos have three different as well as important consequences in the BSM physics. First, it provides the explanation of tiny neutrino masses via Type-I seesaw mechanism. Second, as the right-handed neutrinos are charged under $U(1)_{B-L}$ gauge group, it gives rise to the much needed annihilation mode for $B-L$ charged but SM gauge singlet scalar DM candidate. The $s$-channel annihilation occurs via the $B-L$ symmetry breaking scalar which is a SM gauge singlet. Furthermore, the displaced decay of the right-handed neutrinos provide interesting signatures at the LHC and future colliders which can be used to indirectly constrain the mass of the  $B-L$ gauge boson $Z^\prime$.

The requirement of correct DM relic density needs the annihilation of DM pair to the right-handed neutrino pair which decay further to the SM particles. Such decays of right-handed neutrinos are included in the analysis and the impact of the decay effect is prominent in Fig.~\ref{fig:reso}.
 By changing the decay width, the mass gap between the two points in the DM mass axis is completely eliminated and observed relic is satisfied over a $~30$ GeV range near the $B-L$ scalar $S_0$ resonance region. It thus makes the parameter space viable for models. Given the bounds on $m_{Z^\prime}$ from collider experiments, we concentrate on the scenario where $s$-channel annihilation via $B-L$ symmetry breaking scalar is dominant. Existence of such scalar is crucial, not only for generating the $Z^\prime$ boson mass but also for obtaining the correct DM abundance. However, production of such SM singlet scalar is challenging due to the absence of coupling with the quarks. Given the tiny mixing angle with SM Higgs boson and with $m_{S_0}> m_h$, it is difficult to discover such scalar at the LHC.

We study the bound on the mass of $Z^\prime$ from SI DM-nuclei scattering cross-section measured in direct detection experiments. As the scattering cross-section is proportional to $\qd^2$, $\qd$ being the $B-L$ charge of the DM, the limits on $m_{Z^\prime}$ are more stringent with increasing charge of the DM and thus competes with recent $Z^\prime$ searches at the LHC. 

In this paper, we also investigate the production and decays of the right-handed neutrinos at the LHC. The right-handed neutrinos decay into the SM gauge boson, Higgs boson and leptons via Type-I mixing terms and the decay widths are proportional to the Yukawa coupling $y_N^2$. Consequently, the decay widths are much smaller and lead to the displaced decays of the right-handed neutrinos and thus result in displaced charged track of leptons. Signals of this kind are mostly free from SM backgrounds and easy to probe at the LHC. We show a data of 100\,fb$^{-1}$ of integrated luminosity is sufficient to probe $m_{Z'} \sim 5.5$ TeV at the 14\tev LHC.

\section*{Acknowledgments }
R.M. thanks the Korea Institute for Advanced Study, Seoul for the hospitality during the initial part of the project.
P.B. acknowledges The Institute of Mathematical Sciences, Chennai for the visit for part of the duration of the collaboration. This project has received funding from the European Union's Horizon 2020 research and innovation programme under the Marie Sklodowska-Curie Grant Agreement No. 690575.

\end{document}